\documentclass[sigconf]{acmart}

 \usepackage{enumitem}
\usepackage[capitalise,nameinlink]{cleveref}
\AtBeginDocument{%
  }

\setcopyright{acmlicensed}
\copyrightyear{2018}
\acmYear{2018}
\acmDOI{XXXXXXX.XXXXXXX}

\acmConference[XXX]{XXX}{XXX}{XXX}
\acmISBN{978-1-4503-XXXX-X/18/06}

\begin{document}

\title{Sedeve-Kit, a Specification-Driven Development Framework for Building Distributed Systems}

\author{Hua Guo}
\orcid{0009-0003-4410-0636}

\email{hguo@dase.ecnu.edu.cn}
\affiliation{%
  \institution{East China Normal University}
  \city{Shanghai}
  \country{China}}

\author{Yunhong Ji}
\email{jyh_2017@outlook.com}
\affiliation{%
  \city{Shenzhen}
  \state{Guangdong}
  \country{China}
}

\author{Xuan Zhou}
\email{xzhou@dase.ecnu.edu.cn}
\affiliation{%
  \institution{East China Normal University}
  \city{Shanghai}
  \country{China}}

\renewcommand{\shortauthors}{Guo et al.}

\begin{abstract}
Developing distributed systems presents significant challenges, primarily due to the complexity introduced by non-deterministic concurrency and faults. To address these, we propose a specification-driven development framework. 
Our method encompasses three key stages.
The first stage defines system specifications and invariants using TLA${^+}$. It allows us to perform model checking on the algorithm's correctness and generate test cases for subsequent development phases.
In the second stage, based on the established specifications, we write code to ensure consistency and accuracy in the implementation.
Finally, after completing the coding process, we rigorously test the system using the test cases generated in the initial stage. 
This process ensures system quality by maintaining a strong connection between the abstract design and the concrete implementation through continuous verification.

Screencast URL \footnote{https://www.youtube.com/watch?v=IKqJ6UX1q2o}.

\end{abstract}



\makeatletter
\def\tok@scan#1{%
  \ifx#1\relax
    \let\tok@next\relax
  \else
    \edef\my@list{\my@list#1}%
    \let\tok@next\tok@scan
  \fi
  \tok@next
}
\newcommand{\@strip}[2]{%
  \def\my@list{}\tok@scan#2\relax\let#1\my@list}
\newcommand{\Cite}[1]{\@strip\@args{#1}\cite\@args}
\makeatother

\maketitle

\section{Introduction}

Developing a distributed system is inherently complex, and ensuring its correctness and reliability is even more challenging. Traditional development and quality assurance methods often fail to guarantee the quality of distributed systems. Formal verification and specification are valuable for ensuring the mathematical correctness of systems design \Cite{formal_spec:journals/toplas/ClarkeES86, SpecifyingSystems:books/aw/Lamport2002}. However, there is often a gap between design and implementation, as formal methods verify the design rather than the final implementation.
It is impractical to model all possible program state details in the specification at the design level. At the implementation level, a system's behavior often involves numerous lines of code, with system state updates scattered throughout the codebase. Without careful engineering design,   the final implementation may deviate from the original design as software evolves and iterates, leading to quality defects.

Since formal methods ensure the correctness of the design, we can implement the program according to the specifications and verify it using the specification's model state space to ensure it refines the specification. Based on this idea,  we developed \emph{\textbf{S}p\textbf{e}cification-\textbf{D}riv\textbf{e}n De\textbf{ve}lopment Tool\textbf{kit} (sedeve-kit)} framework, which ensures software quality throughout the entire software engineering cycle via specification-driven development (SDD). We specify algorithms using TLA${^+}$~\Cite{SpecifyingSystems:books/aw/Lamport2002} and verify them with the model checker. Through correct formal specifications, sedeve-kit automatically generates test cases, which are then used to validate the final implementation. These test cases exhaustively cover the state space corresponding to the design, eliminating the need for developers to design test cases manually and significantly reducing their workload.
 
This paper makes the following contributions:
(1) We establish sedeve-kit, a specification-driven development framework that optimizes distributed system development.
Our source code is available for open access on GitHub\footnote{\label{kit_repo}https://github.com/scuptio/sedeve-kit }.
(2) Our method ensures the correctness of design and implementation, maintaining compliance with the design specifications.
(3) We develop a general abstraction to describe distributed systems using I/O automata \Cite{IOAutomata:conf/podc/LynchT87} and TLA${^+}$, and create tools and libraries to map this abstraction to its implementation.
(4) We create a testbed that is capable of running deterministic tests guided and controlled by the specification, liberating developers from manual tasks such as writing test cases, preparing test case data, and simulation.

\section{Preliminaries}

\subsection{I/O Automata}

We use the Input/Output automata  (I/O automata or automata) abstraction to formalize the system. 
An automata is a model represented as a simple state machine that transitions from one state to another, and each transition is called an action. 
 
An I/O automata $A$ has the following components:

\begin{itemize}[leftmargin=*]
  \item 
  
  $\textbf{sig(A)}$: An  \emph{action signature} $S$ describes the I/O automata by three disjoint sets of actions, the input actions $in(S)$, the output actions $out(S)$  and the internal actions $int(S)$.
   
  \item
  
  $\textbf{state(A)}$: A finite \emph{set of states}; a special state is the \emph{initial states}, a nonempty subset of $state(A)$ which is denoted as ${\mathit{start}(A)}$.  

  \item 
  
  $\textbf{{trans}(A)}$: A \emph{state transition relation}; for a non-empty subset $s$ of $\mathit{state}(A)$ and an action $\pi$ ,  there is a nonempty state set $s'$, which is a subset of $\mathit{state}(A)$, and the transition $(s, \pi, s') \in \mathit{trans}(A)$.
  
  \item
  $\textbf{{task}(A)}$: A \emph{task partition} represents a combination of action sequences grouped to form a single task.
\end{itemize}

We then defined a \emph{trace} of $A$, is a finite sequence: 

  $\{ s_0, \pi_1, s_1, \pi_2, s_2, ...,  \pi_r, s_r \}$,
  
in which, for $i \in \{0,1,...,r\}$,  $s_i \in \mathit{state}(A)$, 
$(s_i, \pi_{i+1}, s_{i+1}) \in \mathit{trans}(A)$, and $s_0$ is ${\mathit{start}(A)}$.
The state transition graph of an I/O automata produces a collection of \emph{trace} sets, which are used to generate a set of test cases.

\subsection{TLA$^+$}

TLA$^+$ (Temporal Logic of Actions) \Cite{SpecifyingSystems:books/aw/Lamport2002} is a formal specification language.  TLA$^+$ allows software engineers and system designers to precisely describe the behavior and properties of a system using mathematical notation.
TLA$^+$ specifications can be checked using the TLC model checker, which exhaustively explores all possible system behaviors to ensure that specified properties hold under all circumstances. This rigorous verification process helps detect potential errors, uncover corner cases, and improve the overall reliability of system designs.

\section{The Design of Sedeve-Kit}

\subsection{The Components of Sedeve-Kit}

\begin{figure}
  \centering
  \includegraphics[width=\linewidth]{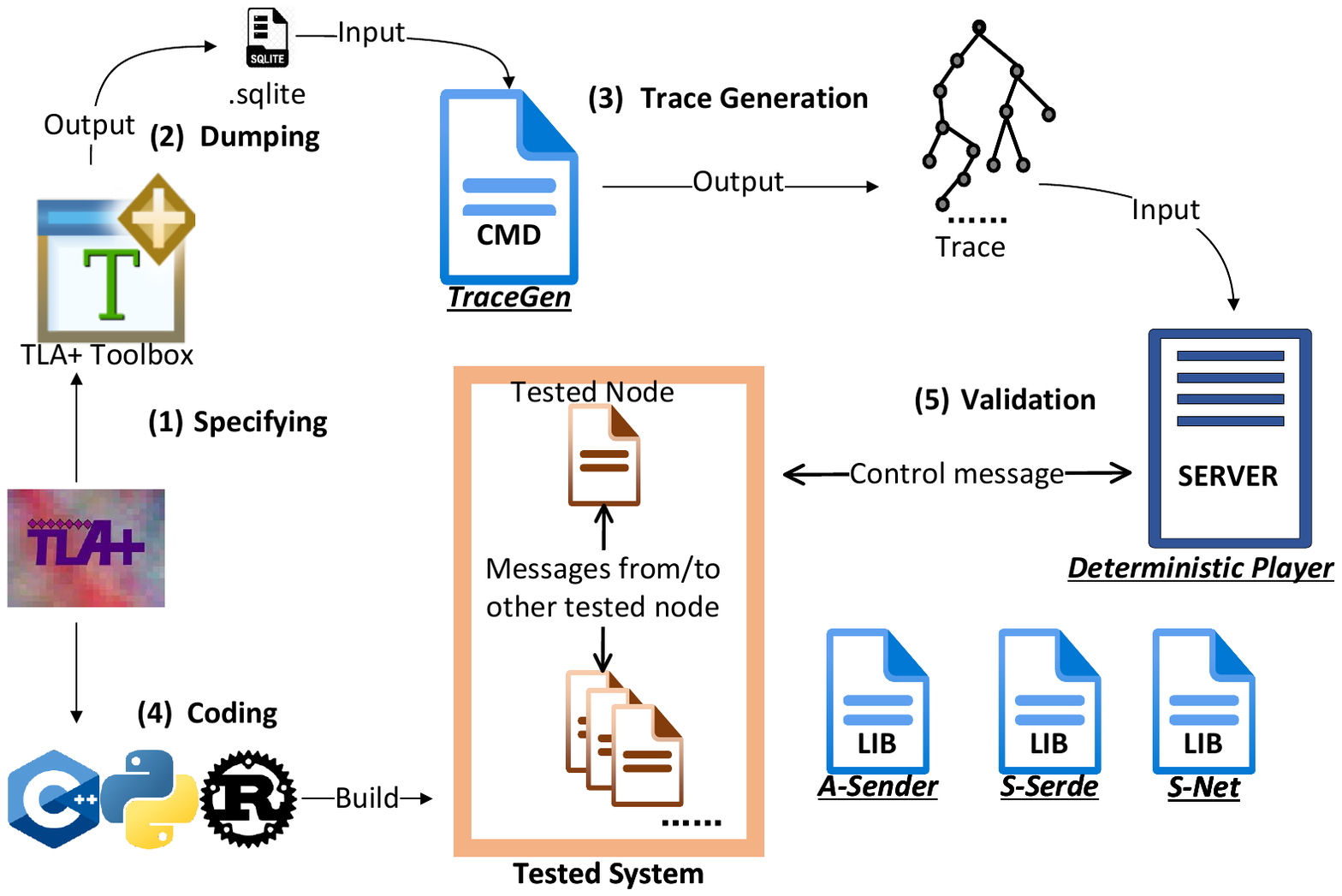}
  \caption{The components of sedeve-kit, in which the icons with bold, italicized, and underlined text titles indicate the tools and libraries provided by the kit.}
  \Description{The components of sedeve-kit}
  \label{fig:architecture}
\end{figure}

Sedeve-kit contains a collection of libraries and command-line tools for the developer to facilitate SDD.
\Cref{fig:architecture} shows the architecture of the \emph{sedeve-kit}.
The following is the components of the \emph{sedeve-kit}.

\begin{itemize}[leftmargin=*]
  \item \emph{TraceGen}, used to generate \emph{trace} set and test cases. It receives input from the SQLite database that stores the TLC model's states, which are retrieved by overwriting TLC operators \Cite{TLAPLUSToolbox:journals/corr/abs-1912-10633} developed by us\footnote{https://github.com/scuptio/SedeveModules}.
  \emph{TraceGen} generate all traces using depth-first search \Cite{DFS:books/daglib/0023376}. The detailed algorithm can be found in the project pages ${^\text{\ref{kit_repo}}}$.

  \item \emph{D-Player} ({Deterministic Player}), controlling to run actions as the trace order.
   
  \item \emph{A-Sender} (Action Sender), a library used by tested systems to send/receive control messages from \emph{D-Player} to run deterministic testing.
  
  \item \emph{S-Serde} (Serialization), a library used to marshal or unmarshal network and control messages.
  
  \item \emph{S-Net}, a library wrapping the network interface.

\end{itemize}

\subsection{The Workflow of SDD Using Sedeve-Kit}

To begin, we initiate the design of a system at an abstract level. 
We use TLA$^+$ to specify the I/O automata (\textbf{(1)} in \Cref{fig:architecture}). 
Next, we run a model checker on this specification and ensure they pass the model-checking process, guaranteeing correctness and dumping the model's states (\textbf{(2)} in \Cref{fig:architecture}).
Then, we use \emph{TraceGen} to generate a trace set (\textbf{(3)} in \Cref{fig:architecture}).
We then write the code based on the specification and incorporate \emph{action anchor macros (AAM)}(which would be discussed in \Cref{sec:action_anchor_macro}) at appropriate source locations (\textbf{(4)} in \Cref{fig:architecture}).  
During the compilation for testing, these macros establish a communication channel with the \emph{D-Player} and reorder actions in a predetermined order generated by the specification.
When the program is compiled for release, the macros would left empty and have no effect.
Finally, we validate the system using our \emph{D-Player} and the \emph{trace} test case(trace) set (\textbf{(5)} in \Cref{fig:architecture}). 
We repeat this procedure until all the test cases pass successfully.
 This process is iterative and can evolve continuously. 

\subsection{Modeling I/O Automata With TLA$^+$}

The system developer must identify the system events that should be defined as actions of I/O automata and their respective action types (input, output, or internal) before building the systems. 
However, not all system events need to be modeled as actions, 
we do not model deterministic behaviors.
Deterministic behavior refers to the property that the program produces the same output when given the same input.  
Most sequential code operating within the same thread is typically deterministic,  and it has easily understandable behavior. 
Non-deterministic behavior, stemming from failures, concurrency, random and network messages, especially asynchronous messages that can be lost, arrive out of order, or duplicated, is difficult to test manually. 
Such behaviors can lead to unexpected results and are challenging to reproduce. 
To model a system is mostly about modeling its non-deterministic behaviors and taking them as deterministic input/output/internal actions of the automata.
We use TLA$^+$ to specify the non-deterministic behaviors and the automata's components. 
The subsection will explore how these components can be modeled using TLA$^+$.

\subsubsection{Action signature and state}
\label{sec:action_signature}
Automata defines three action types, \emph{input}, \emph{internal}, \emph{output} actions.
We introduce an auxiliary variable, ${\_\_action\_\_}$, for every TLA$^+$  specification.  
The auxiliary variable keeps the action signatures and the previous states of the action. 
When a state transitions to the next state, the auxiliary variable ${\_\_action\_\_}$ is updated to capture essential states to express the action signature. 
This information includes the action type (it can be an input, output, or internal), the action name, and the corresponding task (node ID in the context of modeling a distributed system) that yields,  and any additional relevant contexts associated with the action. 
The states are also saved by ${\_\_action\_\_}$.
In the specification, we use TLA$^+$ to write an initial predicate condition to compute an initial state.
A \emph{Next} operator generates the next state from a previous state.

\subsubsection{State transition relation}
We write TLA$^+$'s \emph{Next} operators to describe how each variable changes at each step.
 The \emph{Next} operator defines a condition predicate for how the transit can happen.
 If the current state satisfies the predicate condition, then the state transitions to the next state.
 The transition would be an element of the global transition relation set.
 We get the final state transition relation set by adding all transitions.

\subsubsection{Task partition}
The task partition of a distributed system can be considered a partition of the state transition set on one node in the system. 
In TLA${^+}$, we identify a node by its node ID.

\begin{figure}
  \centering
  \includegraphics[width=\linewidth]{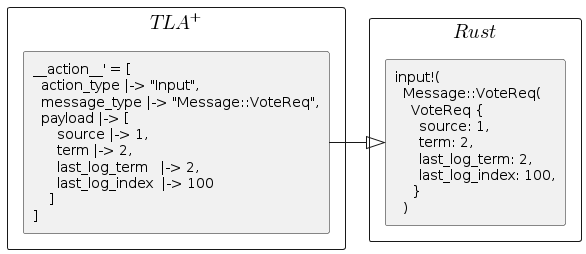}
  \caption{Mapping TLA$^+$ action to anchor macro in Rust code}
  \label{fig:tla_to_rs}
\end{figure} 
\subsection{Mapping the System Behaviour to the Referenced  I/O Automata Action}
\label{sec:action_anchor_macro}

In the TLA$^+$ specification, we use a ${\_\_action\_\_}$ variable to store the necessary states and signatures of I/O automata. 
When programming and being guided by the specification, the developer maps the system behaviors to the referenced action.
In the system implementation level, we provide several \emph{action anchor macros (AAM)} to map the ${\_\_action\_\_}$ variable of TLA$^+$. These macros include:
  \begin{itemize}
    \item     ${\mathit{Input}(I)}$, used when there is an input action ${I}$;
    \item ${\mathit{Output}(O)}$, used when there is an output action ${O}$;
    \item ${\mathit{BeginInternal}(T)}$, used when initiate internal action ${T}$;

    \item ${\mathit{EndInternal}(T)}$, used when finalize internal action ${T}$.
  \end{itemize}

The developers insert these macros into the system's source code based on the system's specifications.
 Most of the actions are about handling messages from network channels. 
When running deterministic testing, these macros make sense, overwrite the network channel, and communicate to \emph{D-Player} to conduct the validation.
\emph{D-Player} reads trace dumped from the model using 
\emph{TraceGen} and enforces executing a predefined action order.
\emph{TraceGen} can read actions from ${\_\_action\_\_}$ to construct the \emph{trace} of all possible trace of action sequences.

\Cref{fig:tla_to_rs} illustrates the mapping from the RequestVote TLA$^+$ action to the Rust source code in the Raft protocol \Cite{Raft:conf/usenix/OngaroO14}.

\subsection{Testing Controlled by D-Player to Validate The
 System}

\label{sec:sub:simulating}

We developed \emph{D-Player} and \emph{A-Sender} to control running tests and simulations with predefined action orders.
The \emph{A-Sender} library is used by the system to send action control messages to the \emph{D-Player} and read the response to control the order of actions and check states.
The \emph{D-Player} receives action control messages from the \emph{A-Sender} library and reorders them according to the \emph{trace} order defined by reading from the SQLite database generated previously as \textbf{Step} \textbf{(3)} in \Cref{fig:architecture}.
Actions are controlled to run one by one in the order specified by the trace. 
After \emph{D-Player} receives an \emph{A-Sender}'s action control request, it would compare the action with the current action of \emph{trace}'s step.
If the action control request received is not the one expected for the current step, the RPC (Remote Procedure Call) request will be blocked and wait until it is the step's turn for that action.
When the action's turn comes, the \emph{D-Player} responds with a message to let the invoking \emph{A-Sender} pass and answers the current states of the automata to let the system check its state consistency with the model.
\emph{A-Sender} and \emph{D-Player} communicate through \emph{S-Serde} and \emph{S-Net} libraries developed by us.
The \emph{A-Sender} library translates \emph{AAM} (which, like ${input!(...)}$ in source code, into RPC requests sent to the \emph{D-Player} and reorder actions (for testing) or left empty (for releasing).

If an inconsistency occurs, \emph{D-Player} would report the errors.
Formally, suppose there is a trace $T$, ${T = \{s_0, \pi_1, s_1, \pi_2, s_2, ... \pi_n, s_n\}}$, in which $\pi_i$ is the $i$th  action of the $T$, and ${s_i}$ is the system's state after runing ${\pi_i}$.
The player processes each action ${\pi_i}$ in $T$. After executing action ${\pi_i}$, the system verifies its state by asserting that its current state matches the expected state  ${s_i}$. 
If the system receives an action ${\pi_i}$, then it yields an action ${\pi'_{i+1}}$ that does not match the expected following action ${\pi_{i+1}}$, the player will trigger a timeout to report the inconsistency. 
 The developer could quickly identify how the error occurred by debugging the error trace and examining each action within the trace.

\begin{figure}
  \centering
  \includegraphics[width=\linewidth]{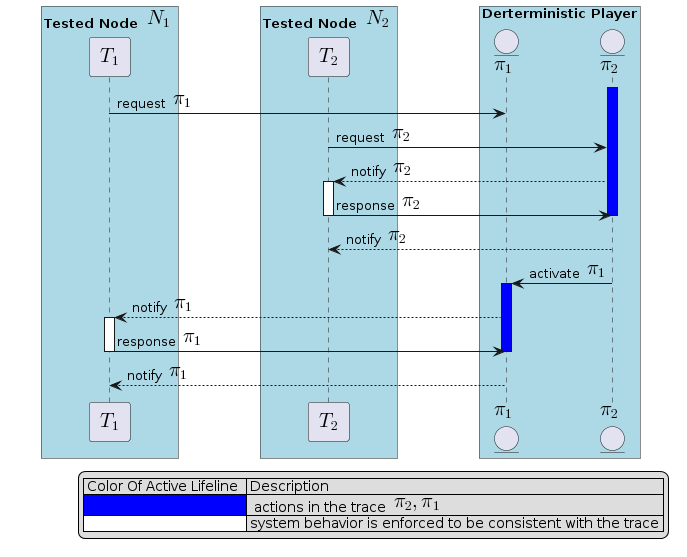}
  \caption{
    The system has two nodes $N_1$ and $N_2$. The two nodes run two tasks $T_1$ and $T_2$, which run two actions $\pi_1$ and $\pi_2$ concurrently. 
    The \emph{D-Player} reads an \emph{trace} $\pi_1$, $\pi_2$. 
    Then, the player enforces the order of $\pi_1$ and $\pi_2$ as the \emph{trace}'s order.
  }
  \label{fig:reorder_actions}

\end{figure}

\Cref{fig:reorder_actions} is an example sequence diagram showing the message flow between the tested system and the \emph{D-Player} and how to enforce the order of action sequence. 
Two nodes are being tested in the figure. 
Node ${N_1}$ executes action ${\pi_1}$ in task ${T_1}$, while node ${N_2}$ executes action ${\pi_2}$ in task ${T_2}$. 
The predefined \emph{trace} order of the test case is ${\pi_2}$ before ${\pi_1}$, but task ${T_1}$ starts action ${\pi_1}$ before task ${T_2}$ starts action ${\pi_2}$.
To ensure the consistent order of actions with the \emph{trace} order, node ${N_1}$ and ${N_2}$ communicate with the \emph{D-Player} through \emph{A-Sender} library to request the reordering of actions. 
In this case, the task ${T_1}$ would be blocked and wait until task ${T_2}$ finishes executing ${\pi_2}$.
The white timelines in node ${N_1}$ and ${N_2}$ indicate that ${\pi_2}$ executes before ${\pi_1}$. 
This order is consistent with the blue lifelines shown in the \emph{D-Player}.
This figure demonstrates how the \emph{A-Sender} and \emph{D-Player} enforce the predefined order of actions that are meant to be concurrent, executing them in the \emph{trace} order.
We implemented the \emph{A-Sender} library and \emph{D-Player} with Rust. We also provide a C binding wrapper to support more languages.

\section{Comparison With State-Of-The-Art Works}

There are some state-of-the-art works similar to ours. They could be characterized by two types: Formal Verification Framework(FVF) 
\Cite{IronFleet:conf/sosp/HawblitzelHKLPR15, Verdi:conf/pldi/WilcoxWPTWEA15, Chapar:conf/popl/LesaniBC16, Lightweight:conf/sosp/BornholtJACKMSS21} and Model-Based Testing(MBT) \Cite{eXtreme:journals/pvldb/SchvimerDH20, Mocket:conf/eurosys/WangDGWW023}.
Like FVF, our work enforces implementation as a specification refinement. 
Unlike FVF, we minimize the effort required to comply with the specification and proof.
FVF usually requires a lot of non-trivial work.
For example, Verdi \Cite{Verdi:conf/pldi/WilcoxWPTWEA15} uses 12511 SLOC (source lines of code) Coq to specify the basic Raft protocol(no log compaction and membership management) and 36925 SLOC to prove it.
We use 3038 SLOC TLA${^+}$ (including invariants) to specify the entire functional Raft \footnote{\label{raft} https://github.com/scuptio/scupt-raft }, including log compaction and membership management.
Like MBT methods, unlike FVF, our work uses testing cases to confirm the specification and does not require much manual work. 
Our work is not designed to find bugs in systems like MBT \Cite{Mocket:conf/eurosys/WangDGWW023}. 
Applying for a legacy system may be our limitation, but it can be improved by introducing an adaptation layer on top of our \emph{A-Sender} libraries. 
MBT \Cite{eXtreme:journals/pvldb/SchvimerDH20, Mocket:conf/eurosys/WangDGWW023} could not find some concurrency bugs if the behavior of the system violated the specification. 
Such missing bugs caused by deviating from the specification cannot exist in our work because our implementation is a specification refinement; the wrong schedule against the specification cannot pass the testing.
 The spectrum of our work lies between FVF and MBT.

Our advantages include:

1. Applicability to any concurrency and distributed systems, failure models, any programming languages, and network environments instills developers' confidence in correctness and software quality.

2. Minimal effort is required for compliance with the specification and proof, and easy integration into software development cycles facilitates the closure of the design and implementation phases.

3. It can be easily integrated into the continuous integration(CI) / continuous delivery(CD) process to ensure the quality of the entire software life cycle.

Our disadvantages are as follows: 

1. When it applies to legacy systems, the developer must build an additional adapter layer(serialization and network libraries) and specify system behaviors;

2. Assurance of correctness only within the context of the specification, i.e., the design space, without providing a wholly correct and bug-free system as system build by FVF \Cite{Verdi:conf/pldi/WilcoxWPTWEA15, IronFleet:conf/sosp/HawblitzelHKLPR15,SEL4:conf/sosp/KleinEHACDEEKNSTW09}.

\section{Conclusion}

We developed sedeve-kit, an SDD framework for building a distributed system.
It can seamlessly integrate into continuous development and deployment processes, enhancing early defect discovery, increasing productivity, and enabling faster release cycles. The framework guarantees software quality by shifting defect detection to the leftmost side of the software development lifecycle.
\bibliographystyle{ACM-Reference-Format}
\bibliography{sample-base}

\end{document}